\documentclass[11pt,a4paper]{article}

\usepackage{arxiv} 

\usepackage[T1]{fontenc}
\usepackage[utf8]{inputenc}
\usepackage{authblk}

\author[1]{Alexander Partin, \thanks{Correspondence: apartin@anl.gov}}
\author[1]{Thomas Brettin}
\author[1]{Yitan Zhu}
\author[2]{James M. Dolezal}
\author[2]{Sara Kochanny}
\author[2]{Alexander T. Pearson}
\author[1]{Maulik Shukla}
\author[3]{Yvonne A. Evrard}
\author[4]{James H. Doroshow}
\author[1,5]{Rick L. Stevens}

\affil[1]{Computing, Environment and Life Sciences, Argonne National Laboratory, Lemont, IL, USA}
\affil[2]{Department of Medicine, Section of Hematology/Oncology, University of Chicago Medical Center, Chicago, IL, USA}
\affil[3]{Frederick National Laboratory for Cancer Research, Leidos Biomedical Research, Inc. Frederick, MD, USA}
\affil[4]{Division of Cancer Therapeutics and Diagnosis, National Cancer Institute, Bethesda, MD, USA}
\affil[5]{Department of Computer Science, The University of Chicago, Chicago, IL, USA}

\title{Data augmentation and multimodal learning for predicting drug response in patient-derived xenografts from gene expressions and histology images}

\usepackage[utf8]{inputenc} 
\usepackage[T1]{fontenc}    
\usepackage{hyperref}       
\usepackage{url}            
\usepackage{booktabs}       
\usepackage{amsfonts}       
\usepackage{nicefrac}       
\usepackage{microtype}      
\usepackage{caption}
\usepackage{subcaption}
\usepackage{comment}
\usepackage{lipsum}
\usepackage[dvipsnames]{xcolor}
\usepackage{multirow}
\usepackage{soul}

\usepackage{graphicx}
\usepackage{afterpage}
\usepackage{array}
\usepackage{booktabs}
\usepackage{amssymb}
\usepackage{amsbsy}

\usepackage{siunitx} 

\graphicspath{ {./figures/} }

\begin{document}
\maketitle

\begin{abstract}
\textbf{Background.}
Prediction of drug response is a critical research area in precision oncology and has been previously explored with large drug screening studies of cancer cell lines (CCLs). Patient-derived xenografts (PDXs) are an appealing platform for preclinical drug studies because the in vivo environment of PDXs helps preserve tumor heterogeneity and usually better mimics drug response of patients with cancer compared to CCLs.

\textbf{Methods.}
We investigate multimodal neural network (MM-Net) and data augmentation for drug response prediction in PDXs. The MM-Net learns to predict response using drug descriptors, gene expressions (GE), and histology whole-slide images (WSIs) where the multi-modality refers to tumor features only. The MM-Net uses late integration where separate subnetworks are used to encode the different input feature types before concatenation and prediction layers. Median tumor volume per treatment group is assessed relative to the control group to create a binary variable representing response.
We explore whether the integration of WSIs with GE improves predictions as compared with models that use GE alone. We use two methods to address the limited number of response values in the dataset: 1) homogenize drug representations which allows to combine single-drug and drug-pairs treatments into a single dataset, 2) augment drug-pair samples by switching the order of drug features which doubles the sample size of all drug-pair samples. These methods enable us to combine single-drug and drug-pair treatments, allowing us to train multimodal and unimodal neural networks (NNs) without changing architectures or the dataset.

\textbf{Results.}
Prediction performance of three unimodal NNs which use GE are compared to assess the contribution of data augmentation methods. NN that uses the full dataset which includes the original and the augmented drug-pair treatments as well as single-drug treatments significantly outperforms NNs (p-values < 0.01) that ignore either the augmented drug-pairs or the single-drug treatments. In assessing the contribution of multimodal learning based on the MCC metric, MM-Net statistically significantly outperforms all the baselines (p-values < 0.05).

\textbf{Conclusion.}
Our results show that data augmentation and integration of histology images with GE can improve prediction performance of drug response in PDXs.

\end{abstract}


\section{Introduction} \label{sec:intro}
With recent advancements in applications of artificial intelligence in medicine and biology, predictive modeling has gradually become one of the primary directions in cancer research for analytically predicting the response of tumors to anticancer treatments \cite{ballester_artificial_2021, adam_machine_2020}. In particular, conventional machine learning (ML) and deep learning (DL) methods have been widely investigated for building computational drug response prediction models for cancer cell lines (CCLs) with large datasets of omics profiles \cite{sharifi-noghabi_drug_2021}. The complex heterogeneities of cancer that occur within and between tumors present a major obstacle to successful discovery of robust biomarkers and therapies \cite{fisher_cancer_2013, seoane_challenge_2014}. Patient-derived tumor xenografts (PDXs) are a contemporary biological model that is created by grafting cancerous tissue, obtained from human tumor specimens, into immunodeficient mice. The in vivo environment of PDXs helps preserve tumor heterogeneity as compared to in vitro CCLs, and therefore, is presumed to better mimic the response of human patients with certain cancer types. PDXs continue gain reputation for studying cancer and investigating drug response in preclinical drug studies \cite{Evrard2020, goto_patient-derived_2020, yoshida_applications_2020}.

Predicting the response of tumors to drug treatments with accurate and robust computational models provides a modern approach for identifying top candidates for preclinical drug screening experiments or personalized cancer treatments. A variety of ML and DL approaches have been explored with high-throughput drug screens with CCLs \cite{chiu_deep_2020, zhu_ensemble_2020}. Alternatively, our literature search retrieved only two publications that have used only PDX data to train prediction models for drug response \cite{nguyen_predicting_2021, kim_pdxgem_2020}. Both studies used the Novartis PDX data (NIBR PDXE), which were generated using a 1x1x1 experimental design \cite{migliardi_inhibition_2012, gao_high-throughput_2015}, where each drug was tested against each patient PDX model using only one entumored mouse per model. Nguyen et al. used an optimal model complexity (OMC) strategy with random forests to build drug response models for 26 treatment-cancer type combinations \cite{nguyen_predicting_2021}. They considered three genomic feature types in their analyses, including gene expressions (GE), copy-number alterations (CNAs), and single-nucleotide variants (SNVs). While considering a single feature type at a time, they used OMC to determine an optimal subset of features to obtain the best performing model for each treatment-cancer type pair. They showed that for the majority of cases, models developed with OMC outperform models that used all the available features. In another study, Kim et al. proposed PDXGEM, a pipeline that identifies biomarkers predictive of drug response in PDX and then uses the identified markers to train prediction models \cite{kim_pdxgem_2020}. To identify predictive genes based on GE and drug response, the pipeline utilizes a strategy similar to co-expression extrapolation (COXEN) \cite{smith_coxen_2010, zhu_enhanced_2020}, and consequently selects the genes whose co-expression patterns are best preserved between PDXs and patient tumors. They trained prediction models using random forests for six treatment-tumor type combinations and then predicted response in patients.

A primary challenge in modeling drug response with PDXs is the limited availability of drug response data. The sample size of PDXs is usually orders of magnitude smaller than the analogous CCL datasets. It has been shown that increasing the amount of training samples improves generalization performance of supervised learning models in vision and text applications \cite{hestness_deep_2017, rosenfeld_constructive_2020}, as well as drug response models in CCLs \cite{xia_cross-study_2021, partin_learning_2021}. Collecting PDX response data, either through experiments or integration of multiple datasets, carries considerable technical and financial challenges. Alternatively, instead of directly increasing the sample size, the volume of data can be expanded by representing each sample with multiple feature types. Multimodal architectures that integrate genomic and histology images have been shown to improve prognosis prediction of patients with cancer as compared with unimodal architectures that learn only from a single data modality (i.e., feature type) \cite{Mobadersany2018, cheerla_deep_2019, chen_pathomic_2019}. Another possible direction to address the limited sample size is data augmentation. Augmentation techniques have been extensively explored with image and text data, but not much with drug response. While augmenting images has become a common practice, tabular datasets such as omics profiles lack standardized augmentation methods.

In this study, we investigate two approaches for predicting drug response in PDX, including multimodal learning and data augmentation. We explore a multimodal neural network (MM-Net) that learns to predict drug response in PDXs using GE and histology whole-slide images (WSIs), two feature types representing cancer tissue, and molecular descriptors that represent drugs. The multimodal architecture is designed to take four feature sets as inputs: (1) GE, (2) histology images, and (3,4) molecular descriptors of a pair of drugs. We benchmark the prediction performance of MM-Net against three baselines: (1) NN trained with drug descriptors and GE, (2) NN trained with drug descriptors and WSIs, and (3) LightGBM model \cite{ke_lightgbm_2017} trained with drug descriptors and GE. With multimodal learning, our goal is to explore whether the integration of histology images with GE improves the prediction performance as compared with models that use GE features alone. For data augmentation, we homogenize the drug representation of single-drug and drug-pair treatments in order to combine them into a single dataset. Moreover, we introduce an augmentation method that doubles the sample size of all drug-pair treatments. The proposed augmentations allow us to combine single-drug and drug-pair treatments to train MM-Net and the baselines without changing the architectures and the dataset. We explore the contribution of augmented data for improving the prediction of drug response.

This paper provides unique contributions compared with existing works that train drug response models with PDX data \cite{nguyen_predicting_2021, kim_pdxgem_2020}. First, we build general drug response models for PDXs across multiple cancer types and drug treatments. Alternatively, prediction models in \cite{nguyen_predicting_2021, kim_pdxgem_2020} are built for specific combinations of cancer type and drug treatment. Thus, our study targets a more challenging task. Second, we utilize PDX histology images with multimodal architecture which has not yet been studied for drug response prediction in PDX. Our study presents a framework for integrating image data with genomic measurements and drug chemical structure for predicting treatment effect. Third, we combine multiple treatments into a single dataset by homogenizing single-drug and drug-pair treatments and utilize drug features for training models. This provides an advantage over existing works which built prediction models for unique combinations of drugs and cancer types, and therefore, disregard drug features when model training. Furthermore, we propose an augmentation method for drug-pairs that doubles the sample size of the drug-pair treatments in the dataset. Fourth, existing studies built prediction models using the PDXE drug screening data which were generated using a 1x1x1 experimental design, i.e. one mouse per model per treatment. In contrast, we utilized the PDMR dataset where treatment response is measured by comparing a group of treated mice to a group of untreated mice. The group approach allows to assess the variability of response across mice and might be considered as more reliable in capturing tumor heterogeneity.

\section{Methods} \label{sec:methods}

\subsection{Data}

\subsubsection{Experimental Design of Drug Efficacy in PDX} \label{sec:pdx_exp_design}
\begin{figure}[h]
    \centering
    \includegraphics[width=0.9\textwidth]{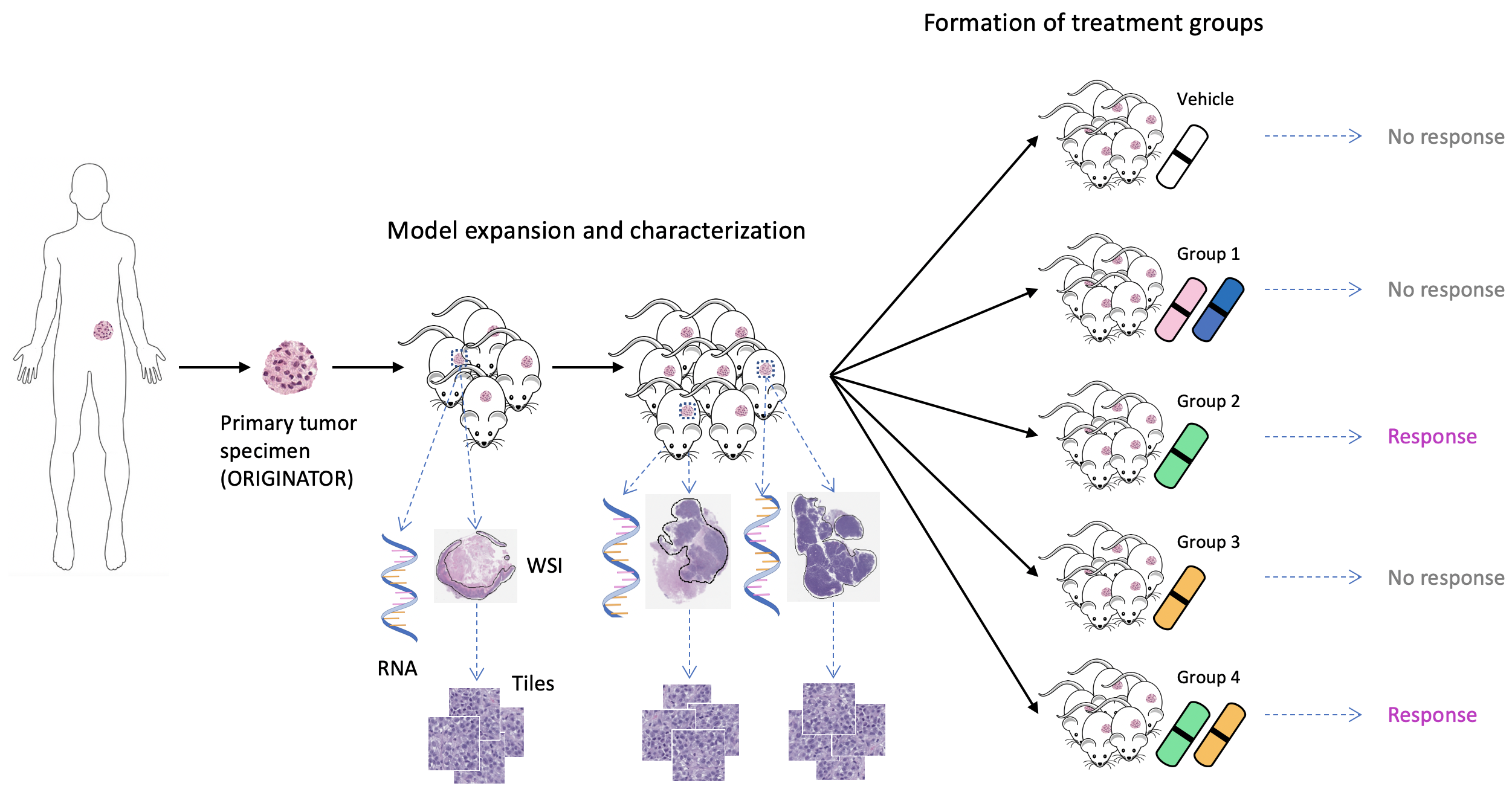}
    \caption{Expansion of tumor tissue from the source specimen (ORIGINATOR) to mice across multiple passages. Mice originated from the same specimen are divided into a control group and multiple treatment groups. Tumors from certain mice were histologically and molecularly profiled resulting in whole-slide images and omics profiles.}
    \label{fig:exp_design}
\end{figure}

We used unpublished PDX drug response data from the NCI Patient-Derived Models Repository (PDMR; \url{http://pdmr.cancer.gov}). The NCI PDMR performs histopathology assessment, whole-exome sequencing and RNA-Seq analysis of a subset of tumors from each PDX model to establish baseline histology and omic characterization for each model. To date, over six hundred unique PDX models have been characterized and data made available through the public website.  Baseline pathology and omic characterization from 487 models were used for this analysis. The efficacy of drug treatments in PDMR is measured through controlled group experiments. Fig. \ref{fig:exp_design} illustrates the process of obtaining primary tumor specimen from a patient, engrafting tumor tissue into PDX models, performing baseline characterization, expanding tumor tissue over multiple passages within a lineage, and then using the expanded tumors in drug treatment experiments. A total of 96 PDX models from 89 unique patients were used for the experiments. The tumors are grown subcutaneously in NOD.Cg-Prkdcscid Il2rgtm1Wjl/SzJ (NSG) host mice and staged to an approximate tumor weight of 200 mm$^3$ for the drug studies. The control group is treated with a vehicle only (vehicle is a solution that delivers the drug in the treated groups). The preclinical dataset includes twelve single-drug and 36 drug-pair treatment arms. Median tumor volume over time for each vehicle or treatment group is used for response assessment. The GE profiles and WSIs were aggregated and preprocessed for the downstream ML and DL analysis.

\subsubsection{Drug Response in PDX} \label{sec:drug_resp}
\begin{figure}[h]
    \centering
    \includegraphics[width=0.7\textwidth]{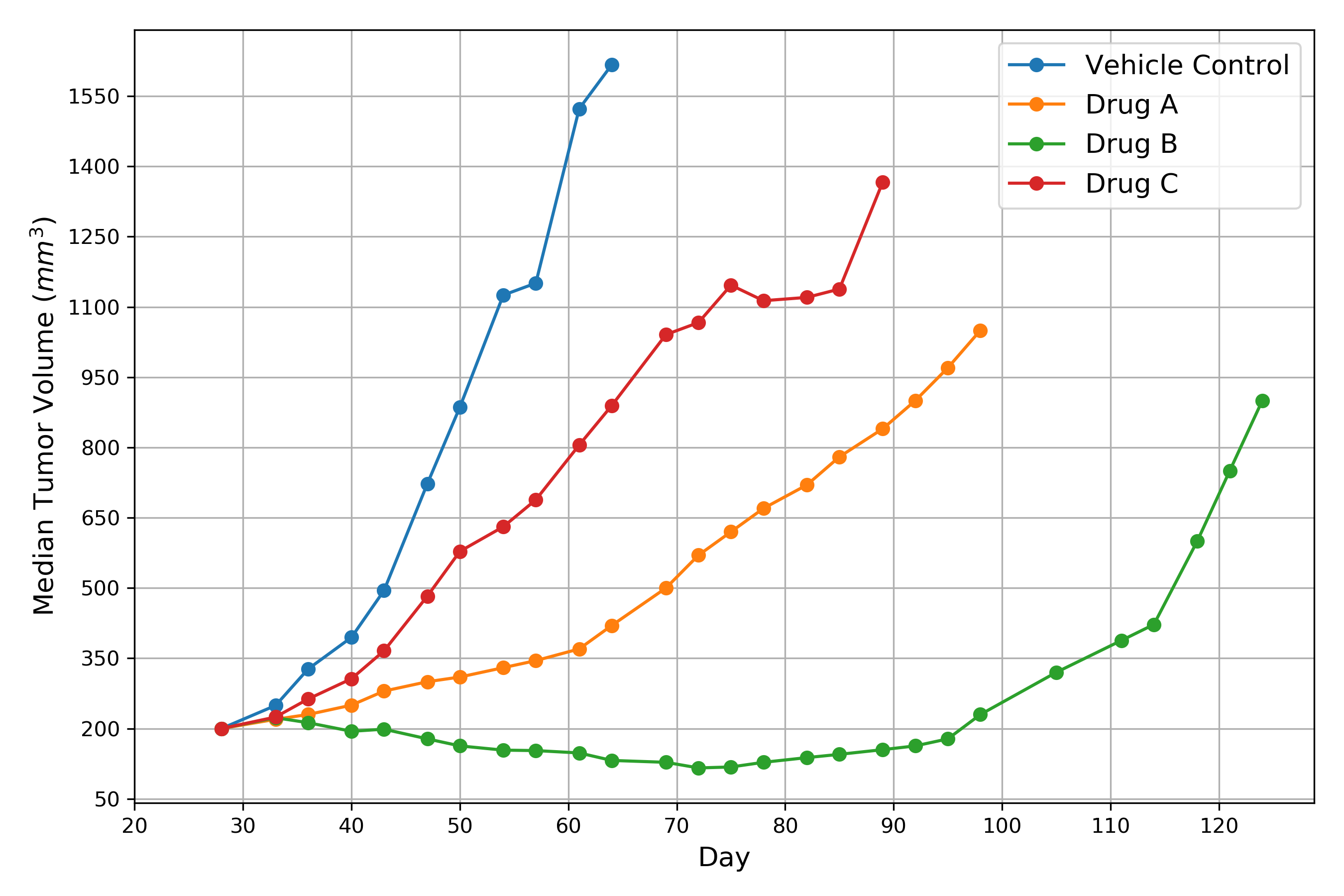}
    \caption{Representative tumor growth curves for vehicle control and drugs three drugs. Label of response assigned to Drug B as tumor achieved regression and non-response the remaining drugs.}
    \label{fig:pdx_growth_curves}
\end{figure}

The growth of tumor volume over time represents the response of PDX tumors to drug treatments, as shown in Fig. \ref{fig:pdx_growth_curves}. The group approach intends to capture the variability of PDX drug response across mice of the same lineage \cite{mer_integrative_2019}. Median tumor volume per treatment group is assessed relative to the control group to create a binary variable representing response. Specifically, for each drug treatment experiment, a single experienced preclinical study analyst assessed the curves of median tumor volume over time for each treatment arm, and assigned "1" for response (regression of at least 10\% from staging for more than one consecutive time point at any point during the study) and "0" for non-response. In essence, a modified RECIST score for regression versus no regression was used to label the response.  Thus, a single best response value was assigned for each treatment arm.

\subsubsection{Data Generation} \label{sec:data_gen}
Three feature types were used for model training, including drug descriptors, GE, and histology image tiles.

\textbf{Gene Expressions.}
The transcriptomic data of PDXs were generated using RNA-Seq. We converted the values into TPM (transcripts per kilobase million), transformed with \textit{log2}, and finally standardized each gene to have a zero mean and a unit standard deviation. For building drug response models, we selected 942 landmark genes discovered by the Library of Integrated Network-Based Cellular Signatures (LINCS) project \cite{Subramanian2017}, which have been shown to well-represent transcriptomic changes.

\textbf{Drug Descriptors.}
We used the Dragon software package (version 7.0) to calculate numerical descriptors of drug molecular structure. The software calculates various types of molecular descriptors, such as atom types, estimations of molecular properties, topological and geometrical descriptors, functional groups and fragment counts, and drug-like indices. A total of 1,993 descriptors were used for the analysis after removing descriptors with missing values. We standardized the descriptor values across drugs to have a zero mean and a unit standard deviation.

\textbf{Histology Images.}
During PDX model expansion, entumored mice were sacrificed between 1000-2000 mm$^3$ for collection of tumors for representative model characterization including histopathological examination. Hematoxylin and Eosin (H\&E) stained pathology slides were digitized into WSIs using an Aperio AT2 digital whole slide scanner (Leica Biosystems) at 20x magnification. A board-certified pathologist from Frederick National Laboratory of Cancer Research reviewed the slides to ensure the PDX models were consistent with the original patient diagnosis. Tumor regions of interest (ROIs) were annotated within the image slides using QuPath \cite{bankhead_qupath_2017} by a single University of Chicago pathologist.

\begin{figure}[h]
    \centering
    \includegraphics[width=0.8\textwidth]{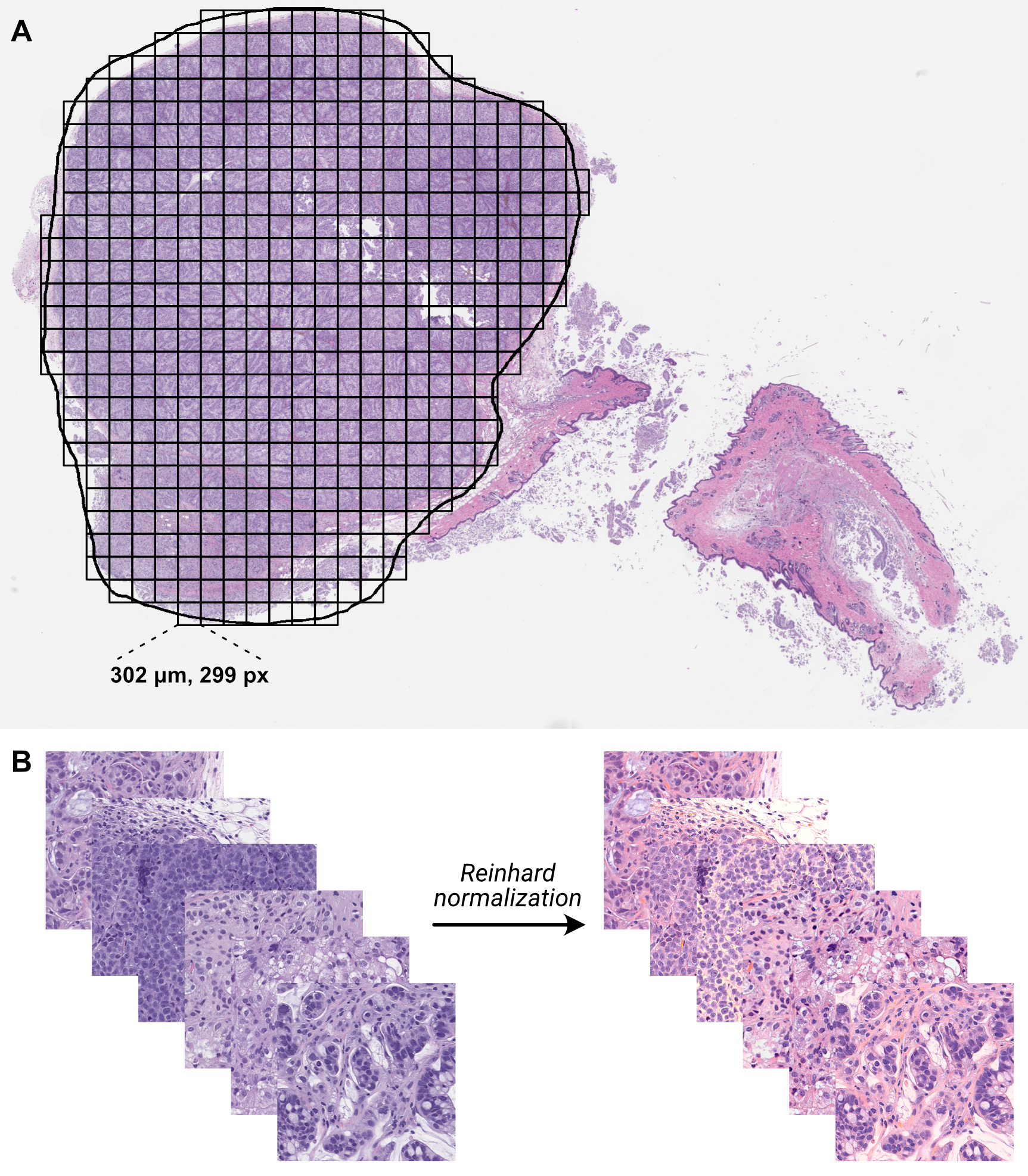}
    \caption{Whole-slide histology image processing. A) Whole-slide images were annotated with region of interest (ROI) outlines, and image tiles were extracted from within ROIs in a grid-wise fashion. B) Extracted non-background tiles underwent digital stain normalization using the Reinhard method \cite{reinhard_color_2001}.}
    \label{fig:wsi_preproc}
\end{figure}

Whole slide images were processed into individual tiles using the Slideflow software package \cite{dolezal_deep_2021, dolezal_jamesdolezalslideflow_2021}, as shown in Fig. \ref{fig:wsi_preproc}. Image tiles were extracted from within annotated ROIs in a grid pattern at 302 $\mu$m by 302 $\mu$m with no overlap, then downsampled to 299 pixels by 299 pixels, resulting in an effective optical magnification of 10x. Background tiles were removed with grayscale filtering, where each tile is converted to the HSV color space and removed if more than 60\% of its pixels have a hue value of less than 0.05. Image tiles then underwent digital stain normalization using the Reinhard method \cite{reinhard_color_2001} and were subsequently standardized to give each image a mean of zero with a variance of one.

\subsubsection{Constructing PDX Drug Response Dataset} \label{sec:ml_dataset}
\begin{figure}[h]
    \centering
    \includegraphics[width=0.8\textwidth]{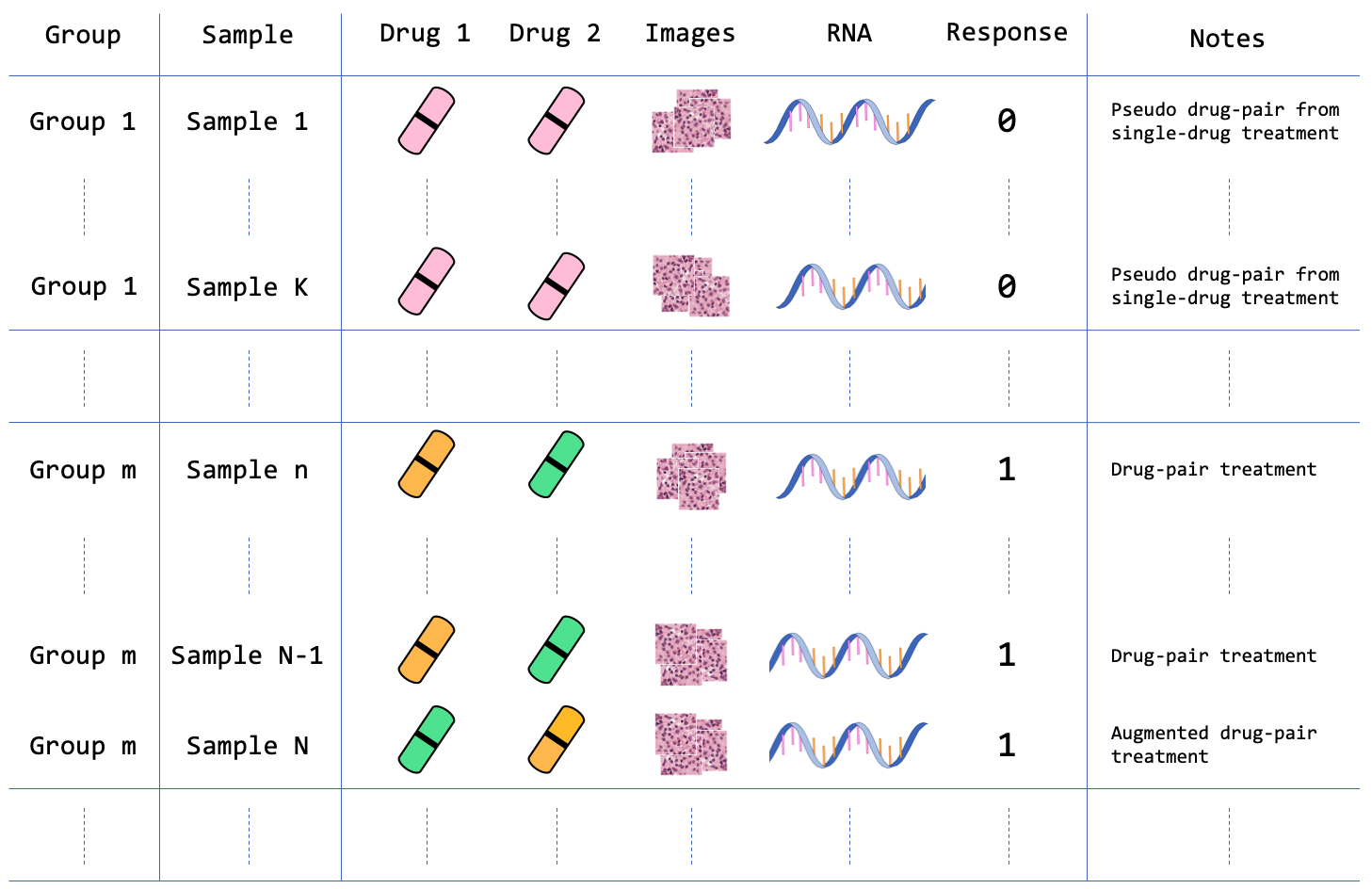}
    \caption{Data arrangement of the PDX drug response dataset. The dataset contains 959 treatment groups after homogenizing and augmenting the drug experiments as described in \ref{sec:ml_dataset}. For example, Sample 1 is a single-drug treatment that is structured as a pseudo drug-pair treatment where Drug 1 and Drug 2 features are the same feature vectors; Sample N is an augmented version of Sample N-1, in which the positions of drug feature vectors are switched. Note that each sample contains multiple histology image tiles that were extracted from a large WSI.}
    \label{fig:tidy_df}
\end{figure}

In constructing the drug response dataset, we populated samples from each group experiment with the corresponding response label. Each sample that was molecularly and histologically profiled consists of three feature types and a binary response value. The feature types include drug descriptors, GE, and histology tiles, as illustrated in Fig. \ref{fig:tidy_df}. Table \ref{tab:summary_stats} lists the summary statistics of the dataset.

The PDMR preclinical dataset contains experiments of single-drug treatments and drug pairs. In order to include both single-drug and drug-pair treatments in the dataset, and ensure consistent dimensionality of drug features, we homogenized single-drug treatments by duplicating drug descriptors to form a pseudo drug-pair that includes two identical drug feature vectors. In this case, the samples of single-drug and drug-pair treatments will follow the same input dimensionality for all ML models. Moreover, because switching the position of drug features in drug-pair treatments should not change the drug response, we augmented all drug-pair samples by switching the position of the two drugs while keeping the drug response value unchanged. Such data augmentation doubles the number of drug-pair samples in the dataset.

Following the integration of group samples into the dataset and the augmentation of drug-pair treatments, the drug response dataset contains 6,962 samples. The total number of treatment groups in the dataset is 959 with 917 non-response and 42 response groups. The dataset contains three feature types (modalities): two vectors of drug descriptors (a vector for each drug), GE profile, and histology tiles. Each sample consists of a unique combination of drug descriptors and GE profiles. However, each such sample contains multiple image tiles from a corresponding histology slide. Concretely, each sample consists of a GE profile, vector of descriptors for drug 1 and drug 2, and multiple image tiles as shown in Fig. \ref{fig:tidy_df}. We store the data in TFRecords (TensorFlow file format) which enables efficient data prefetching and loading, and therefore, considerably decreases the training and inference time.

\begin{table}[h]
\caption{Summary of the PDX drug response dataset used for building prediction models.}
   \centering
   \begin{tabular}{l | rrrrr}
       \toprule
       Patients & 89\\
       Primary tumor specimens & 96\\
       Single-drug treatments & 12\\
       Drug-pair treatments (excludes augmented samples) & 36\\
       Treatment groups	& 959\\
       Gene expression profiles	& 487\\
       Histology whole-slide images (WSIs) & 487\\
       Histology image tiles (extracted from WSIs) & 177,468\\
       Single-drug response values & 2,556\\
       Drug-pair response values (includes augmented values) & 4,406\\
       Drug response values & 6,962\\
       \bottomrule
   \end{tabular}
  \label{tab:summary_stats}
\end{table}

\subsubsection{Data Splits} \label{sec:cv_splits}
Data leakage can lead to overly optimistic predictions \cite{kaufman_leakage_2012}. Two primary characteristics of our dataset may lead to leakage if random splitting is used to generate training, validation, and test sets. First, a drug response label is assigned to all the samples in the entire treatment group. To prevent leakage, we make sure that samples from the same treatment group always appear together in one of the training, validation, or test sets. Second, the augmented drug-pair samples represent, in reality, the same experiment, and therefore, are also placed together when generating the splits. With this strategy, we generated 100 data splits for the analysis (10-fold cross-validation repeated ten times with different random seeds), where the tissue features (GE profiles and WSIs) of the same treatment group are kept together and not shared across training, validation, and test sets of each data split.

\subsection{Prediction Models} \label{sec:pred_models}
\begin{figure}[h]
    \centering
    \includegraphics[width=0.8\textwidth]{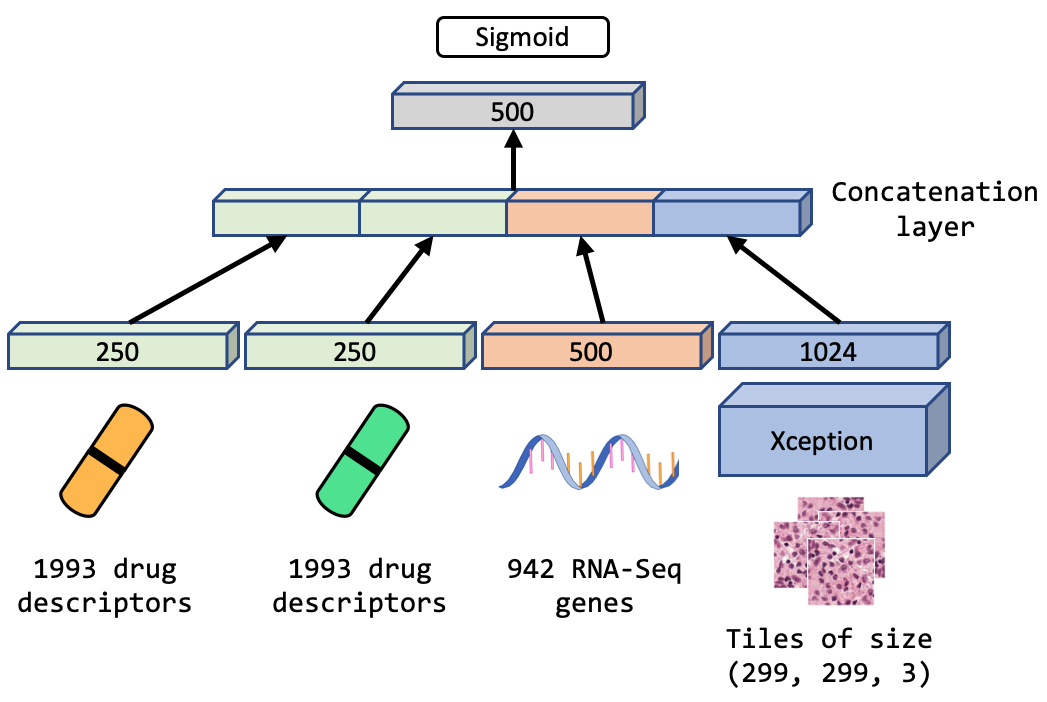}
    \caption{Multimodal neural network (MM-Net).}
    \label{fig:mmNN_arch}
\end{figure}

We explore the performance of MM-Net, shown in Fig. \ref{fig:mmNN_arch}, in predicting the drug response in PDXs. The model takes preprocessed feature sets as inputs, including drug descriptors, GE, and histology tiles, and passes them through subnetworks of layers. The encoded features from the subnetworks are merged via a concatenation layer and propagated to the output for predicting a binary drug response.

Since the dataset is highly redundant in terms of GE and drug features as shown in Fig \ref{fig:tidy_df} and Table \ref{tab:summary_stats} (there are 48 unique drug treatments and 487 unique expression sets), we use a single layer of trainable weights to encode these features with the goal to mitigate overfitting. The image tiles are passed through a subnetwork of convolutional layers of the Xception model \cite{chollet_xception_2017} with weights pre-trained on ImageNet \cite{deng_imagenet_2009}. The output from the convolutional neural network (CNN) is passed through a series of dense layers before being concatenated with the encoded GE and drug descriptor representations.

We compare the performance of MM-Net with three unimodal baselines that use either GE or WSI as tumor features: (1) UME-Net, NN that uses GE, (2) UMH-Net, NN that uses histology tiles, and (3) LGBM, LightGBM that uses GE. Note that all models use drug descriptors.

\subsection{Training and Evaluation} \label{sec:trn_and_eval}
We used a randomized search to obtain a set of hyperparameters (HPs) for UME-Net, including optimizer, learning rate, and layer dimensions that encode GE and drug descriptors. The values of these HPs were also used for MM-Net. A few remaining HPs that are unique to MM-Net were determined in a separate search using the MM-Net architecture. Another randomized search was performed to obtain the HPs for LGBM such as the number of leaves in the decision tree and the number of trees.

To mitigate overfitting, we used the early stopping mechanism in TensorFlow and LightBGM where model trainings terminate automatically if the predictions on a validation set have not been improved for a predefined number of training iterations. The early stopping parameter was set to 10 epochs for all NNs and 100 boosting rounds for the LGBM. All NNs were trained for 400 training epochs which triggered early stopping and ensured model convergence.

Since the dataset is highly imbalanced in terms of drug response distribution, we used a weighted loss function that penalizes more heavily incorrect predictions of the response samples as opposed to the non-response samples. For training MM-Net, we used only 10\% of the image tiles that were available in each WSI, because our preliminary experiments revealed that the prediction performance does not improve if additional tiles are used. The 10\% of the tiles have been drawn at random from each WSI.

For model evaluation, we used three performance metrics for binary classification tasks, including Matthews correlation coefficient (MCC), area under the receiver operating characteristic curve (AUROC), and area under the precision-recall curve (AUPRC). We calculated the metrics based on a test set of each one of the 100 splits. To compute each metric for a given test set, we aggregated all the sample predictions in the test set. In the case of baseline models where each tumor sample is represented by a GE vector, the prediction model generates in a single probability value for each sample. However, in the case of MM-Net where each tumor sample is represented by multiple image tiles in addition to the GE profile, the prediction model generates a single probability value for each image tile which results in multiple predicted probability values for a single sample. To conform with the output of the baseline models, the tile predictions from MM-Net were aggregated via mean to provide a single probability value for each sample. Note that while only 10\% of the available tiles were used for training MM-Net, all tiles in the test set were used to compute predictions and subsequently obtain the performance metrics.

\section{Results} \label{sec:results}

\begin{figure}[ht]
    \centering
    \includegraphics[width=0.8\textwidth]{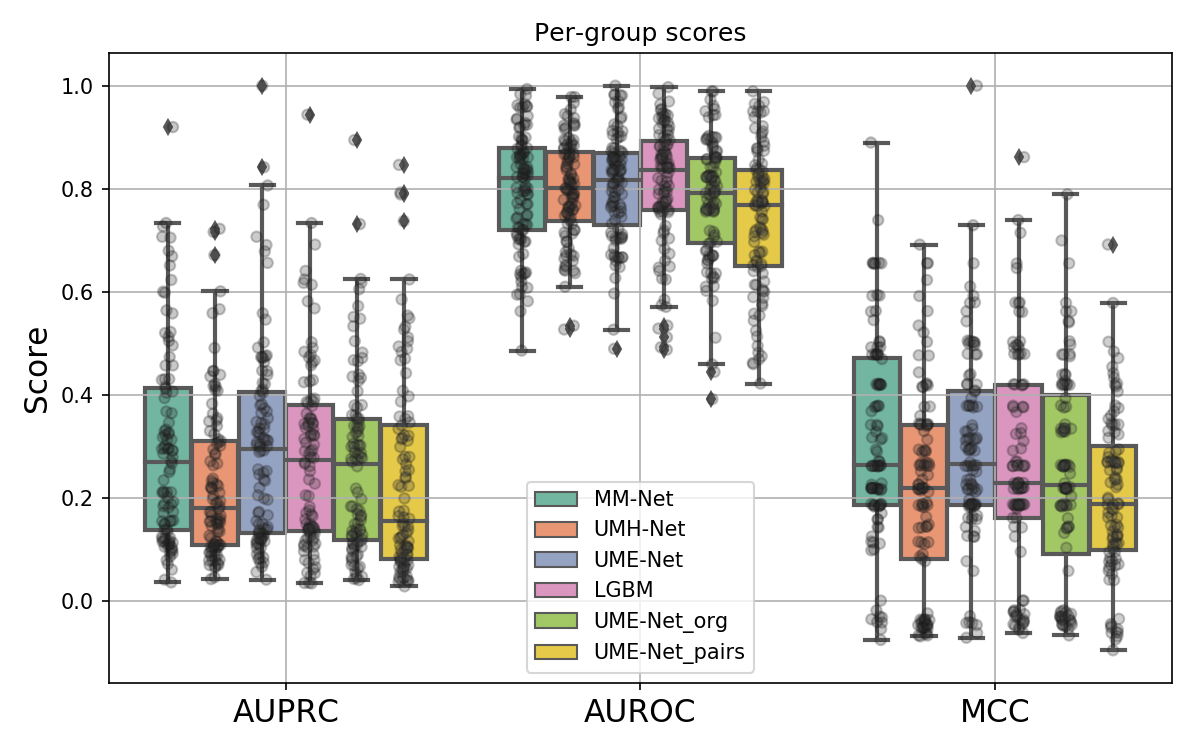}
    \caption{Boxplots showing the distribution of scores for the investigated drug response prediction models. The differences between the models are summarized in Table \ref{tab:table_grp_scores_all}.}
    \label{fig:boxplot_grp_scores_all}
\end{figure}

A total of six prediction models were analyzed as summarized in Table \ref{tab:table_grp_scores_all}. The models differ in terms of the feature sets and the samples that were used for training and validation (binary columns in Table \ref{tab:table_grp_scores_all}). All models were evaluated across the same 100 data splits. Fig. \ref{fig:boxplot_grp_scores_all} shows the performance metrics, including MCC, AUPRC, and AUROC where each data point is a metric value calculated for a given split. The average score of each model was aggregated via mean across the splits for each metric (listed in Table \ref{tab:table_grp_scores_all}).

\begin{table}[ht]
\caption{Performance metrics including MCC, AUPRC, and AUROC are listed for drug response prediction models (UME-Net, UME-Net\textsubscript{org}, UME-Net\textsubscript{pairs}, UMH-Net, MM-Net, and LGBT). As summarized in the binary columns, the differences between the models are in the feature sets (GE, WSI, or both) and the samples. To compute the average score for each metric and model, the predictions were aggregated via mean across the data splits.}
   \centering
   \begin{tabular}{l | rrrrrrrr}
       \toprule
       Model & WSI & GE & Single-drug & Drug-pair & Augmented drug-pairs & MCC & AUPRC & AUROC\\
       \midrule
       UME-Net & -- & v & v & v & v & 0.2958 & \textbf{0.2996} & 0.8047\\
       UME-Net\textsubscript{org} & -- & v & v & v & -- & 0.2391 & 0.2610 & 0.7766\\
       UME-Net\textsubscript{pairs} & -- & v & -- & v & v & 0.2039 & 0.2355 & 0.7423\\
       UMH-Net & v & -- & v & v & v & 0.2124 & 0.2303 & 0.7977\\
       MM-Net & v & v & v & v & v & \textbf{0.3102} & 0.2974 & 0.7978\\
       LGBM & -- & v & v & v & v & 0.2594 & 0.2784 & \textbf{0.8065}\\
       \bottomrule
   \end{tabular}
  \label{tab:table_grp_scores_all}
\end{table}

In constructing the drug response dataset, we used two approaches to increase the number of response values, as described in section \ref{sec:ml_dataset}. We analyzed the effect of these two methods on the prediction performance by comparing three unimodal NNs that were trained with GE and drug descriptors on subsets of the dataset: (1) UME-Net, trained with the full dataset that includes the original and the augmented drug-pair treatments as well as single-drug treatments, (2) UME-Net\textsubscript{pairs}, trained with only drug-pair samples which include the original and the augmented samples, and (3) UME-Net\textsubscript{org}, trained with the original single-drug and drug-pair samples that exclude the augmented drug-pair samples. Fig. \ref{fig:boxplot_grp_scores_all} shows the prediction performance of using the different training subsets across the data splits where each data point is a metric value calculated for a given split. While the distribution of scores across the splits is quite substantial, removing either subset of samples (single-drug samples or augmented drug-pairs) results in a significant decline in prediction performance (p-values < 0.05). In other words, augmentation methods lead to a significant improvement in performance of the NNs when trained with GE and descriptors. Hence, we used the full set of the available training samples for the analyses of multimodal learning.

The MM-Net architecture, shown in Fig. \ref{fig:mmNN_arch}, was compared against three baseline models, including UME-Net, UMH-Net, and LGBM. The performance metrics represent the ability of the models to generalize on a test set of unseen observations. In addition, we computed a paired t-test of the scores across the 100 splits to compare whether there is statistically significant difference between the values. Based on the aggregated MCC, MM-Net statistically significantly outperforms all the baselines (p-values < 0.05) except for UME-Net. When considering the AUPRC and AUROC, MM-Net outperforms UMH-Net but there is no significant difference when comparing MM-Net with UME-Net or LGBM.

\begin{figure*}[ht]
  \centering
  \begin{subfigure}[]{0.45\textwidth}
    \centering
    \includegraphics[width=\textwidth]{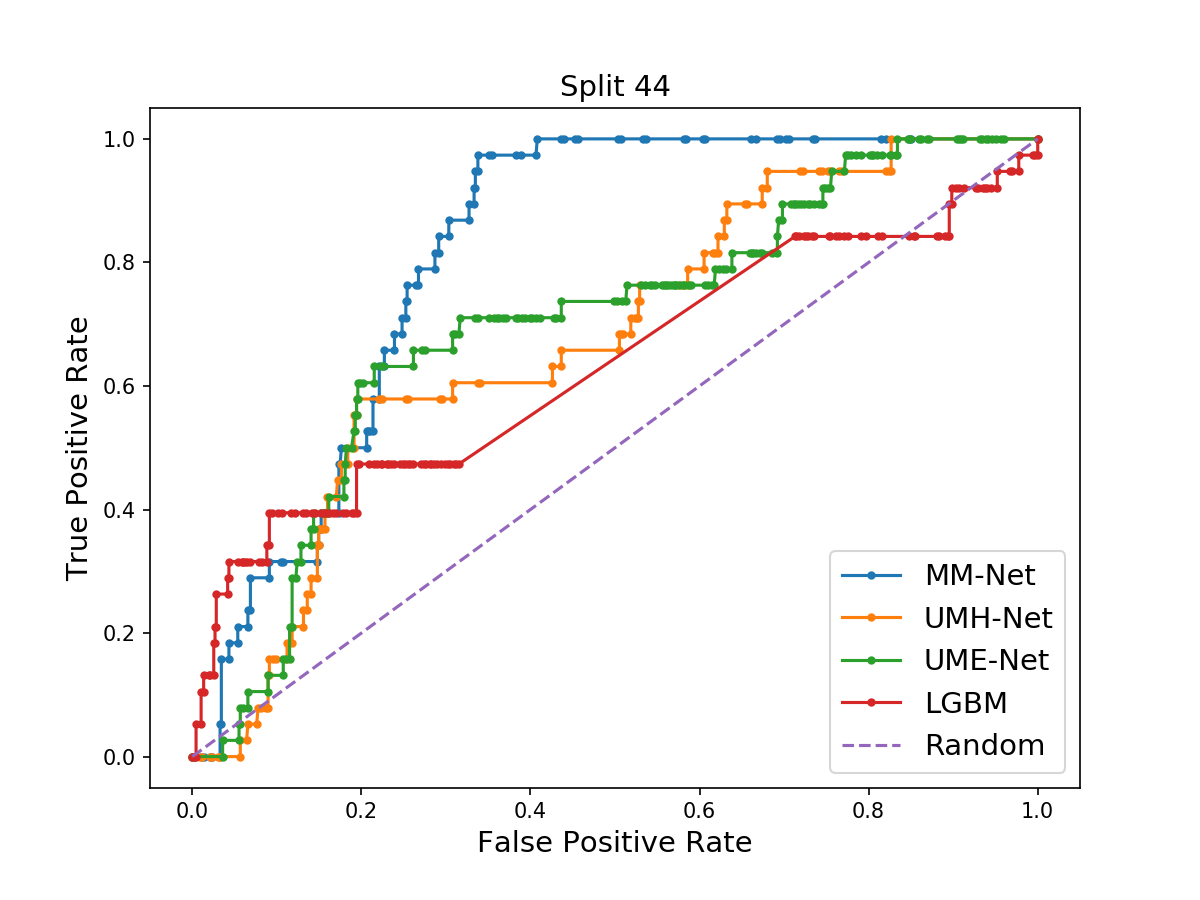}
    \caption{}
    \label{fig:auc_44}
  \end{subfigure}
  \begin{subfigure}[]{0.45\textwidth}
    \centering
    \includegraphics[width=\textwidth]{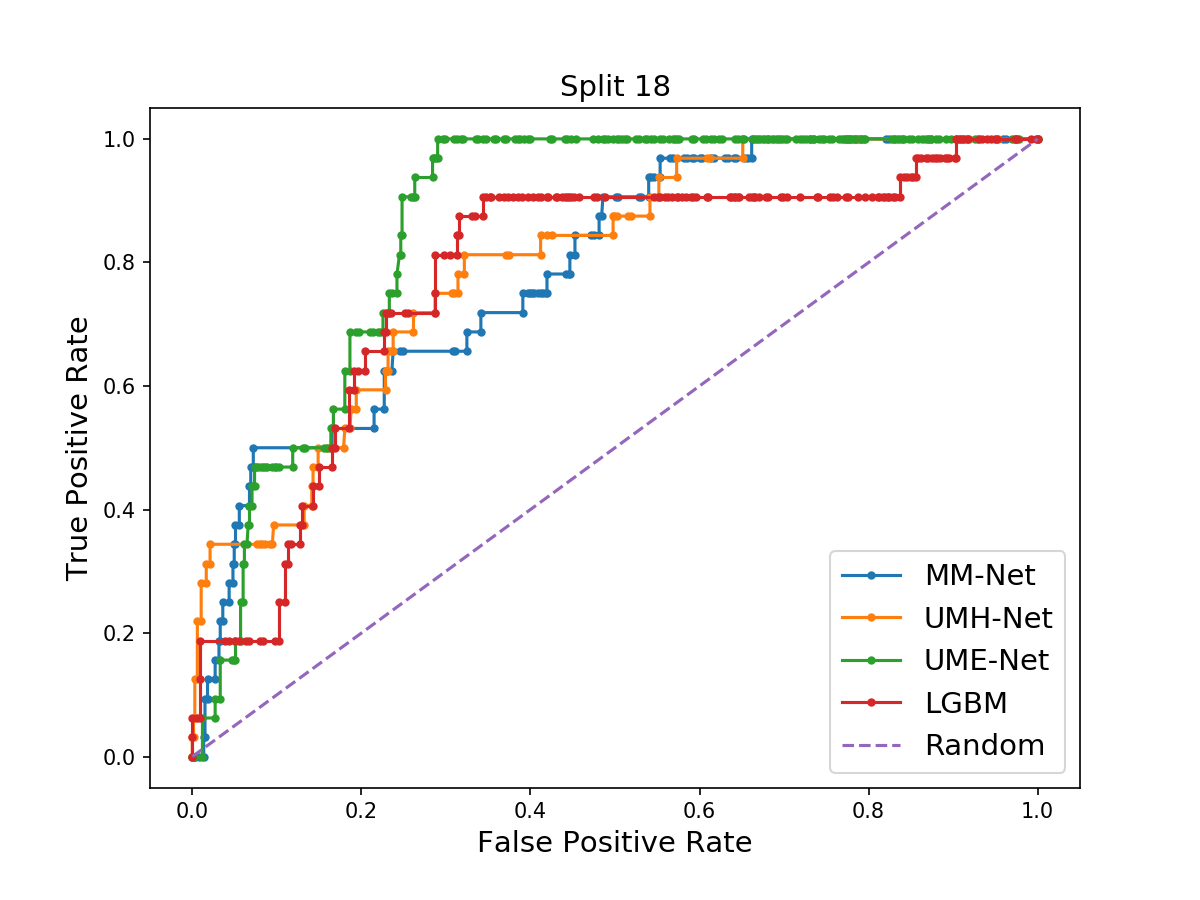}
    \caption{}
    \label{fig:auc_18}
  \end{subfigure}
  \caption{Receiver operating characteristic (ROC) curves for two different data splits illustrating the MM-Net outperforming the baselines on the left plot, and underperforming on the right.}
  \label{fig:roc_plots}
\end{figure*}

We can observe a large spread of scores for all models and metrics. This indicates that for certain data splits, the models exhibit very high generalization performance, while for other splits, the models almost entirely fail to learn a meaningful mapping function for predicting drug response. In practical scenarios, where the goal is to design a highly generalizable model, a careful analysis should be conducted to determine the training and validation sets that adequately represent the test set. In our analysis, however, the objective was to conduct large-scale trainings across multiple data splits and examine the overall capacity of MM-Net in predicting drug response across multiple cancer types and treatments. We observe that for certain dataset splits, MM-Net outperforms the baselines but for other splits it underperforms, as shown in Fig. \ref{fig:roc_plots}. An in-depth investigation is further required to understand in which cases MM-Net trained with WSI significantly improves prediction generalization.

\section{Discussion} \label{sec:discussion}

In this study, we investigated data augmentation methods and a multimodal architecture for predicting drug response in PDXs. We utilized the PDMR drug response dataset of single-drug and drug-pair treatments that were generated in controlled group experiments with PDX models of multiple cancer types. To assess the utility of the proposed methods, we conducted a large-scale analysis by training MM-Net and three baseline models over 100 data splits that contain GE profiles, histology image tiles, and molecular drug descriptors. We demonstrated that data augmentation methods lead to a significant improvement in drug response predictions across all performance metrics (MCC, AUPRC, AUROC). Alternatively, the MM-Net model exhibits statistically significant improvement in prediction performance only when measured by the MCC.

The data splitting strategy and the choice of performance metrics play an important role in the downstream analysis when evaluating the utility of prediction models for practical applications. The dataset size and its diversity in terms of PDX models and treatments allowed us to generate multiple data splits while mitigating data leakage between training, validation, and test sets. Since each split comprises unique GE and histology images, we face a challenging prediction problem as opposed to a situation in which the samples are randomly split. Alternative splitting strategies may involve a careful choice of a single test set with the goal to reduce the distributional shift between training and test sets \cite{stacke_measuring_2021}. Instead of carefully assembling the most representative test set, we chose to conduct a large-scale analysis to assess the empirical range of prediction performance with NNs and LGBM. The results show a large spread of scores across the splits, indicating that for certain data splits the models exhibit high prediction performance, while for other splits, the learning of models fails. When we specifically focus on the performance of MM-Net as compared with the baselines, we discover that in 46 out of the 100 splits, the MM-Net outperformed the UME-Net baseline. This observation implies that for certain data splits, the histology images boost the generalization performance of the prediction model, and therefore, its potential utility in preclinical and clinical settings. A further investigation is required to better understand the cases and data characteristics in which histology images improve response prediction.

Technological progress in digital pathology and high-throughput omic profiling have led researchers to generate big data repositories of histology images and omics data, as well as algorithms to jointly analyze these diverse data types. Several papers have explored multimodal architectures that combine histology images with omics data for predicting survival outcomes of cancer patients. Mobadersany et al. demonstrated that a CNN-based supervised learning model combined with cox regression accurately predicts survival outcomes of glioma patients from histology and mutation data \cite{Mobadersany2018}. Cheerla et al. proposed an unsupervised learning method to learn a low-dimensional representation for each feature type and, consequently, concatenated the learned representations to predict survival outcome of cancer patients \cite{cheerla_deep_2019}. They have also demonstrated on 20 cancer types that a custom dropout layer that randomly drops an entire feature vector improves predictions. Building upon existing works, Chen et al. introduced a supervised architecture for multimodal fusion of histology and omics data to predict patient survival and applied their method to glioma and clear cell renal cell carcinoma patients \cite{chen_pathomic_2019}. The model uses graph convolutional network (GCN) and CNN to encode histology image data and feed-forward network for mutation data. Each encoded feature vector is passed through an attention mechanism and subsequently fused via a Kronecker product. The cox regression is finally used to predict patient survival. While these papers do not consider drug treatments in their analysis, they exploit modern approaches for enhancing predictions of multimodal NNs with histology and omics data and can be further explored for drug response prediction.

Considering the scale of existing PDX datasets, data fusion and augmentation provide promising research directions for enhancing predictive capabilities with PDXs. However, special care should be taken because high-dimensional features sets can often lead to severe overfitting and poor generalization. Presumably to mitigate overfitting, Nguyen et al. \cite{nguyen_predicting_2021} and Kim et al. \cite{kim_pdxgem_2020} used feature selection methods to reduce the dimensionality of PDX data while considering a single omics feature type at a time.  Therefore, multimodal learning exhibits tradeoff between enriching the feature space via multimodal fusion and overfitting. To alleviate this tradeoff, alternative methods can be explored to reduce the feature space while incorporating multiple feature types \cite{zhu_converting_2021}. With the methods proposed in this study and ongoing research into novel augmentation and fusion techniques, PDX pharmaco-omic datasets may become more suitable for modern deep learning techniques and further increase interest for building prediction models to advance precision oncology.

\section*{Funding}
These activities were part of the IMPROVE project under the NCI-DOE Collaboration Program that has been funded in whole or in part with Federal funds from the National Cancer Institute, National Institutes of Health, Task Order No. 75N91019F00134 and from the Department of Energy under Award Number ACO22002-001-00000.

\bibliographystyle{ieeetr}

\bibliography{zotero}  

\end{document}